\documentclass{emulateapj}
\usepackage{apjfonts}

\usepackage{amsmath}

\usepackage{graphics}

\newcommand{\bd}{\begin{displaymath}}
\newcommand{\ed}{\end{displaymath}}

\slugcomment{accepted by ApJ}

\shorttitle{A physical model for active galactic nuclei with
double-peaked broad emission lines}

\shortauthors{Cao \& Wang}

\begin{document}

\title{A physical model for active galactic nuclei with double-peaked broad
emission lines}

\author{Xinwu Cao\altaffilmark{1,3} and Ting-Gui
Wang\altaffilmark{2,3}} \altaffiltext{1}{Shanghai Astronomical
Observatory, Chinese Academy of Sciences, 80 Nandan Road,
Shanghai, 200030, China; cxw@shao.ac.cn} \altaffiltext{2}{Center
for Astrophysics, University of Science and Technology of China,
Hefei 230026, China}\altaffiltext{3}{Joint institute for Galaxy
and Cosmology, Chinese Academy of Sciences, China}

\begin{abstract}

The double-peaked broad emission lines are usually thought to be
linked to accretion disks, however, the local viscous heating 
in the line-emitting disk portion is usually insufficient for
the observed double-peaked broad-line luminosity in most sources.
{It was suggested that the X-ray radiation from an ion-supported
torus in the inner region of the disk can photo-ionize the outer
line-emitting disk region. However, our calculations show that
only a small fraction ($\la$ 2.3 per cent) of the radiation from
the radiatively inefficient accretion flow (RIAF) in the inner
region of the disk can photo-ionize the line-emitting disk
portion, because the solid angle of the outer disk portion
subtended to the inner region of the RIAF is too small.} We
propose a physical model for double-peaked line emitters, in which
only those AGNs with sufficient matter above the disk (slowly moving
jets or outflows) can scatter enough photons radiated from the
inner disk region to the outer line-emitting disk portion
extending from several hundred to more than two thousand
gravitational radii. Our model predicts a power-law $r$-dependent
line emissivity $\epsilon^{{\rm H}\alpha}\propto R_{\rm
d}^{-\beta}$, where $\beta\sim 2.5$, which is consistent with
$\beta\sim 2-3$ required by the model fittings for double-peaked line
profiles. Using a sample of radio-loud AGNs with double-peaked
emission lines, we show that the outer disk regions can be
efficiently illuminated by the photons scattered from slow or
mild relativistic electron-positron jets with $\gamma_{\rm j}\la 2$. It is consistent with the fact that no double-peaked
emission line is present in strong radio quasars with relativistic
jets. For radio-quiet double-peaked line
emitters, slow outflows with Thomson scattering depth $\sim
0.2$ instead of jets can scatter sufficient photons to
(illuminate) the line-emitting regions. This model can therefore
solve the energy budget problem for double-peaked line emitters.

\end{abstract}


\section{Introduction}

Only a small fraction of active glactic nuclei (AGNs) exhibit
double-peaked broad-line profiles, for example, the largest sample
available so far is the 116 double-peaked Balmer line AGNs, which
are found from an initial sample consisting of 5511 broad line
AGNs with $z<0.5$ observed by the Sloan Digital Sky Survey(SDSS)
\citep{s03}. Some previous authors mainly focused on the
double-peaked broad lines in radio-loud(RL) AGNs
\citep*[e.g.,][]{p88,chf89,ch89,eh94,eh03}. A complete survey on
RL AGNs finally resulted in 20 double-peaked line emitters
\citep{eh94,eh03}. It is still a mystery why (only) a small
fraction of AGNs exhibit double-peaked broad-line profiles.

Some different scenarios were suggested for the origin of
double-peaked emission lines, namely, (1) emission from the
accretion disk \citep*[e.g.,][]{chf89,ch89}, (2) emission from a
binary broad-line region (BLR) in a binary massive black hole
system \citep*[e.g.,][]{g83,g88}, (3) emission from the bipolar
outflows \citep*[e.g.,][]{zsb90}, (4) emission from a spherically
symmetric BLR illuminated by an anisotropic ionizing radiation
source \citep*[e.g.,][]{gw96}. The accretion disk model is the
most favorable one among them \citep*[see detailed comparison
between these different scenarios in][and references
therein]{eh03}. The widths of double-peaked lines range from
several thousand to nearly 40000 km~s$^{-1}$ \citep*[e.g.][]{w05}.  In the accretion
disk model, the double-peaked emission lines are radiated from the
disk region between around several hundred gravitational radii to
more than two thousand gravitational radii, and their profiles can
be well fitted by the accretion disks with a power-law line
emissivity \citep*[e.g.,][]{chf89,ch89,eh03}. However, none
of these scenarios can answer why only a small fraction of AGNs
have been detected as double-peaked line emitters. For example, in
the accretion disk model, one may expect to observe double-peaked
lines in most AGNs, as the AGNs with higher inclination angles
have more chances to be observed, until their accretion disks are
obscured by the putative tori \citep*[e.g.,][]{a93}, if the
orientations of AGNs are isotropically distributed in space.
Another difficulty for the disk model is the ``energy budget"
problem, i.e., the local viscously dissipated power in the
line-emitting disk portion is usually insufficient for the
observed double-peaked broad-line luminosity in most sources, and
the temperatures of the line-emitting disk regions are too low to
produce the observed H$_\alpha$ lines
\citep{chf89,ch89,eh94,eh03}. The X-rays from a hot ion-supported
torus in the inner disk region are assumed to irradiate the outer
line-emitting disk region and then to solve the ``energy budget"
problem \citep*[e.g.,][]{ch89}.

{We discuss the illumination of the outer disk regions by the
inner ion-supported tori in \S 2.} In this paper, we suggest a
physical model for these double-peaked line emitters. In this
model, a fraction of photons from the accretion disk are scattered
by the electrons in the jet/outflow back to photo-ionize the outer
line-emitting region, and then to produce the observed emission
lines. We describe our model in \S 3. The cosmological parameters
$\Omega_{\rm M}=0.3$, $\Omega_{\Lambda}=0.7$, and $H_0=70~ {\rm
km~s^{-1}~Mpc^{-1}}$ have been adopted in this work.

\section{Ion supported hot tori in double-peaked line emitters}

The putative hot ion supported tori in double-peaked line emitters
are well modelled by the advection dominated accretion flows
(ADAFs, old version of RIAFs) \citep{ny94,ny95}. The RIAF can
survive only if its accretion rate  $\dot{m}$
($\dot{M}/\dot{M}_{\rm Edd}$) is less than a critical value
$\dot{m}_{\rm crit}$, which is a function of the viscosity
parameter $\alpha$ \citep*[e.g.,][]{ny95}. The three-dimensional
MHD simulations suggest that the viscosity parameter $\alpha$ in
the accretion flows is $\sim 0.1$ \citep{a98}, or $\sim 0.05-0.2$
\citep{hb02}.  For some double-peaked line emitters, their
bolometric luminosities could be as high as one-tenth of the
Eddington value. Their X-ray luminosities $L_{\rm X}^{0.1-2.4\rm
keV}$ could be as high as 10$^{43-44}$ ergs~s$^{-1}$ for the
double-peaked line emitters with black hole masses of 10$^{7-9}\rm
M_\odot$ \citep*[e.g., Table 7 in][and Table 1 in this
paper]{eh03}. The maximal X-ray luminosity for a RIAF surrounding
a massive black hole can be calculated based on the global
solution for the RIAF. Adopting $\dot{m}_{\rm crit}=0.01$,
corresponding to $\alpha\simeq 0.2$, which is a conservative
choice, we can calculate the global structure of the RIAF. For
such a RIAF accreting at the critical rate, its X-ray luminosity
$L_{\rm X}^{0.1-2.4\rm keV}=7.3\times 10^{40}$ ergs~s$^{-1}$ for a
10$^8$ $\rm M_\odot$ black hole \citep*[see][for detailed
description of the calculation on the flow structure]{m00,cao05}. The X-ray luminosity of
the RIAF depends almost linearly on the black hole mass while
$\dot{m}$ is fixed. For some X-ray luminous double-peaked line
emitters, a very high viscosity $\alpha\sim 1$ is required for the
presence of RIAFs in these sources, which is in contradiction with
the MHD simulations.

It is well known that most gravitational energy of accretion
matter is released in a small region close to the black hole,
either for a standard thin disk \citep{ss73} or a RIAF. In Fig.
\ref{fig1}, we calculate the ratio $L_{\rm RIAF}(<R_{\rm
d})/L_{\rm RIAF}^{\rm tot}$ as a function of $R_{\rm d}$ based on
the global RIAF structure \citep[][]{cao05,m00}. It is found that more than 70 per cent radiation
is from the region within 0.3 $R_{\rm d,tr}$, where $R_{\rm d,tr}$
is the radius of the RIAF connecting to the outer standard thin
disk. Assuming  homogeneous distribution of the gases in
$z$-direction of the RIAF, our calculation shows that less than
2.3 per cent radiation from the RIAF in the inner region of the
disk can illuminate the line-emitting disk region with $R_{\rm
d}>R_{\rm d,tr}$, because the solid angle of the outer disk
portion subtended to the inner region of the RIAF is too small.
For more realistic $z$-direction density distribution of the RIAF,
the gases are denser in the midplane of the flow and the resulted
fraction should even be less than this value. Thus, the hot
ion-supported tori are unable to heat the outer line emitting disk
regions efficiently for most double-peaked line emitters, even if the observed luminous X-ray emission
can be attributed to hot ion-supported tori.

\vskip 1.0cm
\figurenum{1}
\centerline{\includegraphics[angle=0,width=7.8cm]{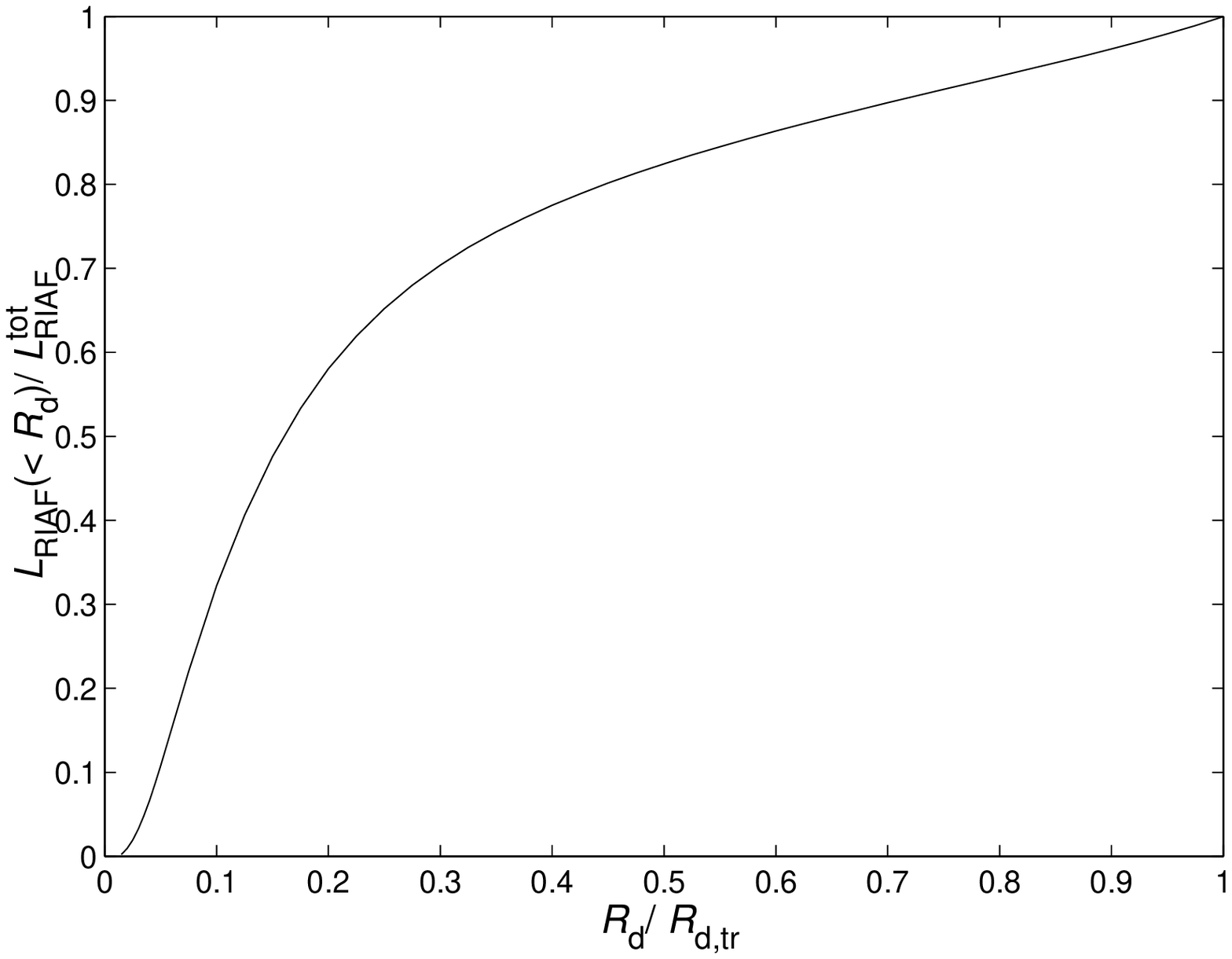}}
\figcaption{The ratio $L_{\rm RIAF}(<R_{\rm
d})/L_{\rm RIAF}^{\rm tot}$ as a function of $R_{\rm d}/R_{\rm
d,tr}$. \label{fig1}}  \centerline{}

\section{The model}

In this paper, we employ the inhomogeneous conical jet model
\citep*[see][for details]{k81}, which can successfully explain
most observational features of radio-loud AGNs
\citep*[e.g.,][]{j98}. For a free jet, its half-opening angle $\phi_{\rm
j}=1/\gamma_{\rm j}$ \citep{bk79,hm86}, where $\gamma_{\rm j}$ is the Lorentz 
factor of the jet. The non-thermal electrons in the jet is described by a power-law
energy distribution,
\begin{equation}
n_{\rm e}(r_{\rm j},\gamma_{\rm e})=n_{\rm e,0}(r_{\rm j})\left({\frac
{\gamma_{\rm e}}{\gamma_{\rm e,min}}} \right)^{-(2\alpha_{\rm
e}+1)},
\end{equation}
from $\gamma_{\rm e,min}$ to $\gamma_{\rm e,max}$ \citep{k81}. 
In K\"onigl's jet model, the jet is assumed to move at a constant 
velocity, which is a good approximation, {though the dynamics 
of electron-positron jets is very complicated \citep*[e.g.,][]{rh98}. 
For simplicity, we adopt a conventional assumption of the jet 
moving at a constant velocity $\beta_{\rm j}c$. }

{It is still unclear for the jet composition, i.e.,
electron-positron or electron-proton \citep*[see][for a recent
review and references therein]{wb04}. For an electron-proton jet,
the lower limit of electron Lorentz factor $\gamma_{\rm e,min}$ is
required to be greater than 100, while $\gamma_{\rm e,min}$ could
be as low as unity for an electron-positron jet
\citep*[e.g.,][]{cf93}. This is consistent with the detection of circular polarization, 
which strongly suggests that the jets are electron-positron plasmas and
have low $\gamma_{\rm e,min}\la 10$ \citep*[e.g.][]{whr98}. The
similar conclusion is arrived from powerful large scale X-ray
jets, if they are interpreted as inverse-Compton scattering of
cosmological microwave background photons in fast jets. The
typical energy of the photons radiated from the standard thin
accretion disk surrounding a massive black hole is $\nu_{\rm d}^*\sim10~{\rm eV}$. The
soft photons from such a standard disk are Compton up-scattered to
the hard X-ray band ($\ga$ 10 keV, or even $\gamma$-ray bands) for
an electron-proton jet, because $\gamma_{\rm e,min}\sim 100$.
These hard X-ray photons can hardly photo-ionize the outer
line-emitting disk region. {If a thermal electron population with similar 
density co-exists with the non-thermal power-law electrons in the relativistic jet, 
the soft X-ray photons scattered by the thermal electrons in the jet can ionize the disk. 
Such thermal plasmas in a relativistic jet may produce significant Doppler shifted emission 
or absorption lines depending on their temperature, 
which seems inconsistent with observations. Thus, we will not consider this 
possibility in this work, though the presence of thermal electrons in the jet cannot be ruled out. } 
The situation is different for
the electron-positron jet, of which $\gamma_{\rm e,min}\sim 1$.
Most Compton scattered photons by the electrons in
electron-positron jets are in soft X-ray bands, which can
efficiently photo-ionize the line-emitting disk region, so 
we will focus on the electron-positron jets hereafter in this work. }

The kinetic luminosity of a relativistic moving electron-positron jet
with Lorentz factor $\gamma_{\rm j}$ is
\begin{displaymath}
L_{\rm kin}=\dot{M}_{\rm jet}(\gamma_{\rm j}-1)c^2
+\dot{M}_{\rm jet}2\alpha_{\rm e}\gamma_{\rm e,min}^{-1}
\left[1-\left( {\frac {\gamma_{\rm e,min}} {\gamma_{\rm e,max}}} \right)
^{2\alpha_{\rm e}}\right]^{-1}
\end{displaymath}
\begin{displaymath}
\times
\int\limits_{\gamma_{\rm e,min}}^{\gamma_{\rm e,max}}\left({\frac {\gamma_{\rm e}}{\gamma_{\rm
e,min}}}\right)^{-(2\alpha_{\rm e}+1)}(\gamma_{\rm e}-1)c^2{\rm d}\gamma_{\rm e}
\end{displaymath}
\begin{equation}
\simeq \dot{M}_{\rm jet}c^2
\left(\gamma_{\rm j}+{\frac {2\alpha_{\rm e}\gamma_{\rm
e,min}} {2\alpha_{\rm e}-1} }-2\right),\label{lkin1}
\end{equation}
where $m_{\rm e}$ is electron rest mass, $\gamma_{\rm
e,max}\gg\gamma_{\rm e,min}$ is assumed, and the mass loss rate of the jet is given by 
\begin{equation}
\dot{M}_{\rm jet}=2m_{\rm e}n_{\rm e}(r_{\rm j})2\pi r^2_{\rm
j}(1-\cos\phi_{\rm j})R_{\rm g}^{2}\gamma_{\rm j}\beta_{\rm j}c. \label{mdotjet}
\end{equation}

For an accretion disk surrounding a black hole, most gravitational
energy is released in the inner region of the disk \citep*[within
several Schwarzschild radii close to the marginal stable orbit
$r_{\rm ms}$,][]{ss73}, where $r_{\rm ms}$ is in units of gravitational  
radius $R_{\rm g}=GM_{\rm bh}/c^2$. For simplicity, we use a ring with radius
$r_{\rm d}=2.25r_{\rm ms}$, where is the main contribution to the
integral disk luminosity \citep{ss73}, to approximate the
accretion disk radiation. The solid angle of the surface of the
jet slice between $r_{\rm j}$ and $r_{\rm j}+{\rm d}r_{\rm j}$
subtended to the accretion disk (ring) with radius $r_{\rm d}$ is
\begin{displaymath}
{\frac {{\rm d}\Omega(r_{\rm j})}{{\rm d}r_{\rm j}}}~~~~~~~~~~~~~~~~~~~~~~~~~~~~~~~~~~~~~~~~~~~~~~~~~~~~~~~~~~~~~~~~~~~~~~~~~~~~~~~~~~~~~~~~~~~~~~
\end{displaymath}
\begin{equation}
={\frac {2\pi
r_{\rm j}r_{\rm d}\sin\phi_{\rm j}\cos\phi_{\rm j}}{[(r_{\rm
d}-r_{\rm j}\sin\phi_{\rm j})^2+r_{\rm j}^2\cos^2\phi_{\rm
j}]\times[(r_{\rm j}-r_{\rm d}\sin\phi_{\rm j})^2+r_{\rm
d}^2\cos^2\phi_{\rm j}]^{1/2}}}, \label{domega}
\end{equation}
where $r_{\rm j}$ is the distance from the apex of the jet. 

The jet is Compton thin in transverse direction at large radii,
while it becomes Compton thick at small radii. We can estimate the
transition radius $r_{\rm j,tr}$ by
\begin{equation}
\tau_{\rm es,tr}(r_{\rm j,tr})=r_{\rm j,tr}R_{\rm g}\phi_{\rm
j}n_{\rm e}(r_{\rm j,tr})\sigma_{\rm T}\sim 1,
\label{tauestr}\end{equation} beyond which the jet becomes Compton
thin in transverse direction for electron scattering. Here the
Thomson cross-section $\sigma_{\rm T}=6.652\times10^{-25}$ cm$^2$.
In the region of the jet with $r_{\rm j}\la r_{\rm j,tr}$, the
photons from the disk are mostly scattered in a thin layer at the
jet surface.

The angle-dependent inverse-Compton scattered photons from the
slice of the jet $r_{\rm j}\rightarrow r_{\rm j}+{\rm d}r_{\rm j}$
measured in the source frame is
\begin{displaymath}
{\frac {{\rm d}N_{\rm Comp}(r_{\rm j},\mu_{\rm L})}{{\rm d}t{\rm d}r_{\rm
j} {\rm d}\Omega_{\rm scat}}}={\frac {f^{\rm IC}L_{\rm disk}}{\pi
h\nu^*_{\rm d}} }{\frac {1-\beta_{\rm j}\cos\phi_{\rm j}
}{\gamma_{\rm j}^2(1-\beta_{\rm j}\mu_{\rm L})^3}}~~~~~~~~~~~~~~~~~~~~~~~~~~~~~~~~~~~~~~~~
\end{displaymath}
\begin{equation}
\times{1\over{2\pi}} {\frac
{r_{\rm j}\cos\phi_{\rm j}}{[(r_{\rm d}-r_{\rm j}\sin\phi_{\rm
j})^2+r_{\rm j}^2\cos^2\phi_{\rm j}]^{1/2}}} {\frac {{\rm
d}\Omega(r_{\rm j})}{{\rm d}r_{\rm j}}},\label{dnscatangthick}
\end{equation}
for the Compton thick part of the jet ($r_{\rm j}\la r_{\rm
j,tr}$), where $\mu_{\rm L}=\cos \theta_{\rm scat}$, and $\theta_{\rm
scat}$ is the angle of the scattered photons measured from the
axis of the jet in the source frame given by
\begin{equation} 
\mu_{\rm L}=-{\frac {r_{\rm j}\cos\phi_{\rm j}}{[r_{\rm
j}^2\cos^2\phi_{\rm j}+(r_{\rm L}-r_{\rm j}\sin\phi_{\rm
j})^2]^{1/2}}}. \label{mu}
\end{equation}
As only the scattered photons with energy $\ge\nu_{\rm ph}^{\rm min}=13.6$ eV can efficiently 
photo-ionize the gases in the line-emitting disk region, the correction 
factor $f^{\rm IC}$ describing the fraction of the scattered photons with energy 
$\ge\nu_{\rm ph}^{\rm min}$ measured in the source rest frame, is given by 
\begin{equation}
f^{\rm IC}=\min\left[\left({\frac {\gamma_{\rm e,min}^\prime}{\gamma_{\rm e,min}} }\right)
^{-2\alpha_{\rm e}},1\right]. \label{fic}
\end{equation}
Only the photons scattered by the electrons with $\gamma_{\rm e}\ge \gamma_{\rm e,min}^\prime$ are 
more energetic than 13.6 eV measured in the source frame, so  
the value of $\gamma_{\rm e,min}^\prime$ is given by
\begin{equation}
\gamma_{\rm e,min}^\prime=\max\left[{\frac {(1-\beta_{\rm j}\mu_{\rm L})^{1/2}
{\nu_{\rm ph}^{\rm min}}^{1/2}}{(1-\beta_{\rm j}\mu_{\rm j}^{\rm s})^{1/2}
{\nu_{\rm d}^{*}}^{1/2}}},1\right], \label{gammaeminp}
\end{equation}
where 
\begin{equation}
\mu_{\rm j}^{\rm s}={\frac {r_{\rm j}-r_{\rm d}\sin\phi_{\rm j}}
{[r_{\rm j}^2\cos^2\phi_{\rm j}+(r_{\rm d}-r_{\rm j}\sin\phi_{\rm j})^2]^{1/2}}}.
\end{equation}

Assuming a line-emitting ring with radii $r_{\rm L}$ extending
from $\xi_1$ to $\xi_2$, we obtain the total photons scattered in
the Compton thick jet slice which can photo-ionize the line-emitting disk
region,
\begin{displaymath}
{\frac {{\rm d}N_{\rm Comp}^{\rm line}(r_{\rm j};r_{\rm j}\le
r_{\rm j,tr})}{{\rm d}t{\rm d}r_{\rm j}
}}=\int\limits_{\mu_{\rm L1}}^{\mu_{\rm L2}}{\frac {f^{\rm IC}L_{\rm disk}}{\pi
h\nu^*_{\rm d}} }{\frac {1-\beta_{\rm j}\cos\phi_{\rm j}
}{\gamma_{\rm j}^2(1-\beta_{\rm j}\mu_{\rm L})^3}}
\end{displaymath}
\begin{displaymath}
\times{\frac {1}{2\pi}}{\frac
{r_{\rm j}\cos\phi_{\rm j}}{[(r_{\rm d}-r_{\rm j}\sin\phi_{\rm
j})^2+r_{\rm j}^2\cos^2\phi_{\rm j}]^{1/2}}}  {\frac {{\rm
d}\Omega(r_{\rm j})}{{\rm d}r_{\rm j}}}2\pi{\rm d}\mu_{\rm L}. 
\end{displaymath}
\begin{displaymath}
=\int\limits_{\mu_{\rm L1}}^{\mu_{\rm L2}}{\frac {f^{\rm IC}L_{\rm disk}}{\pi
h\nu^*_{\rm d}} }{\frac {1-\beta_{\rm j}\cos\phi_{\rm j}
}{\gamma_{\rm j}^2(1-\beta_{\rm j}\mu_{\rm L})^3}}~~~~~~~~~~~~~~~~~~~~~~~~~~~~~~~~~~~~~~~~~~~~~~~~~~~~~~~~~~~~~~
\end{displaymath}
\begin{equation}
\times{\frac {2\pi r^2_{\rm j}r_{\rm d}\sin\phi_{\rm
j}\cos^2\phi_{\rm j}{\rm d}\mu_{\rm L}}{[(r_{\rm d}-r_{\rm j}\sin\phi_{\rm
j})^2+r_{\rm j}^2\cos^2\phi_{\rm j}]^{3/2}\times[(r_{\rm j}-r_{\rm
d}\sin\phi_{\rm j})^2+r_{\rm d}^2\cos^2\phi_{\rm j}]^{1/2}}}.\label{dnscat}
\end{equation}

For the jet part with $r_{\rm j}>r_{\rm j,tr}$, it is Compton thin for
electron scattering. In this case, we have to include radiative
transfer of the incident photons in the jet. Similar to Compton
thick case, we can derive the inverse-Compton scattered photons
from the slice of the jet $r_{\rm j}\rightarrow r_{\rm j}+{\rm
d}r_{\rm j}$ photo-ionizing the line-emitting ring measured in the
source frame as
\begin{displaymath}
{\frac {{\rm d}N^{\rm line}_{\rm Comp}(r_{\rm j};r_{\rm j}>r_{\rm
j,tr})}{{\rm d}t{\rm d}r_{\rm j} }}~~~~~~~~~~~~~~~~~~~~~~~~~~~~~~~~~~~~~~~~~~~~~~~~~~~~~~~~~~~~~~~~~~~~~~~~~
\end{displaymath}
\begin{displaymath}
\simeq\int\limits_{\mu_{\rm L1}}^{\mu_{\rm L2}} 
{\frac {f^{\rm IC}L_{\rm
disk}}{\pi h\nu^*_{\rm d}} }
{\frac {1-\beta_{\rm j}}{\gamma_{\rm j}^2(1-\beta_{\rm j}\mu_{\rm L})^3}}{\frac {1}{4\pi}}
\int\limits_{r^\prime_{\rm j,min}}^{r_{\rm j}}n_{\rm e}(r_{\rm
j})\sigma_{\rm T}R_{\rm g}\exp[-\tau_{\rm es}(r_{\rm
j},r^\prime_{\rm j})]
\end{displaymath}
\begin{equation}
\times{\frac {2\pi {r^{\prime}_{\rm j}}^2 r_{\rm d}\sin\phi_{\rm
j}\cos^2\phi_{\rm j}{\rm d}r^\prime_{\rm
j}2\pi{\rm d}\mu_{\rm L}}{[(r_{\rm d}-r^\prime_{\rm j}\sin\phi_{\rm
j})^2+r^{\prime 2}_{\rm j}\cos^2\phi_{\rm
j}]^{3/2}\times[(r^\prime_{\rm j}-r_{\rm d}\sin\phi_{\rm
j})^2+r_{\rm d}^2\cos^2\phi_{\rm j}]^{1/2}}}, \label{dnscatthin}
\end{equation}
where
\begin{equation} 
\mu_{\rm L}=-{\frac {r_{\rm j}}{(r_{\rm
j}^2+r_{\rm L}^2)^{1/2}}}, \label{mu2}
\end{equation}
and
\begin{equation}
\mu_{\rm L1,2}=-{\frac {r_{\rm j}}{(r_{\rm j}^2+\xi_{1,2}^2)^{1/2}}}.
\label{mu}
\end{equation}
Here the fraction $f^{\rm IC}$ is also given by Eqs. (\ref{fic}) and (\ref{gammaeminp}), while 
$\mu_{\rm j}^{\rm s}$ is given by 
\begin{equation}
\mu_{\rm j}^{\rm s}={\frac {r_{\rm j}}
{(r_{\rm j}^2+r_{\rm d}^2)^{1/2}}}.
\end{equation}
The electron density, $n_{\rm e}(r_{\rm j})\sim r_{\rm j}^{-2}$, for 
a relativistic jet as required by the mass conservation along the jet. 
The optical depth
\begin{displaymath}
\tau_{\rm es}(r_{\rm j},r^\prime_{\rm j})=\cos^{-1}\theta_{\rm
d}(r^\prime_{\rm j})\int\limits_{r_{\rm j}^\prime}^{r_{\rm j}}
n_{\rm e}(r_{\rm
j})\sigma_{\rm T}R_{\rm g}{\rm d}r_{\rm j},
\end{displaymath}
\begin{equation}
={\frac 
{\dot{M}_{\rm jet}\sigma_{\rm T}R_{\rm g}}
{4\pi m_{\rm e}(1-\cos\phi_{\rm j})R_{\rm g}^{2}c\cos\theta_{\rm
d}(r^\prime_{\rm j})\beta_{\rm j}\gamma_{\rm j}}}\left(
{1\over {r_{\rm j}}}-{1\over {r_{\rm j}^\prime}}\right),
\end{equation}
where
\begin{equation}
\cos\theta_{\rm d}(r^\prime_{\rm j})={\frac {r^\prime_{\rm
j}\cos\phi_{\rm j}}{[(r_{\rm d}-r^\prime_{\rm j}\sin\phi_{\rm
j})^2+r^{\prime 2}_{\rm j}\cos^2\phi_{\rm j}]^{1/2}}}.
\end{equation}
The occultation by the Compton thick part of the jet is considered, so 
the lower integral limit of Eq. (\ref{dnscatthin}) is given by
$r^\prime_{\rm j,min}=\max(r_{\rm j,tr},r^{\prime\rm X}_{\rm
j,min})$, and
\begin{displaymath}
r^{\prime\rm X}_{\rm j,min}=[\sin^2\phi_{\rm j}\cos^2\theta^{\rm m}_{\rm
d}(r_{\rm j})+\cos^2\phi_{\rm j}-\cos^2\theta^{\rm m}_{\rm
d}(r_{\rm j})]^{1/2}
\end{displaymath} 
\begin{equation}
\times{\frac {-r_{\rm d}\sin\phi_{\rm
j}\cos^2\theta^{\rm m}_{\rm d}(r_{\rm j})+r_{\rm d}\cos\theta^{\rm
m}_{\rm d}(r_{\rm j})} {\cos^2\phi_{\rm j}-\cos^2\theta^{\rm
m}_{\rm d}(r_{\rm j})}},
\end{equation}
where
\begin{equation}
\cos\theta^{\rm m}_{\rm d}(r_{\rm j})={\frac {r_{\rm
j}\cos\phi_{\rm j}}{[r_{\rm j}^2\cos^2\phi_{\rm j}+(r_{\rm
j}\sin\phi_{\rm j}+r_{\rm d})^2]^{1/2}}}.
\end{equation}

The total photons scattered by the jet that can photo-ionize the
line-emitting disk region is
 \begin{equation}
{\frac {{\rm d}N^{\rm line}_{\rm Comp}}{{\rm d}t }}=
\begin{cases}
\int\limits_{r_{\rm j,min}}^{r_{\rm j,tr}} {\frac {{\rm d}N^{\rm
line}_{\rm Comp}(r_{\rm j};r_{\rm j}\le r_{\rm j,tr})}{{\rm
d}t{\rm d}r_{\rm j} }}{\rm d}r_{\rm j}+\int\limits_{r_{\rm
j,tr}}^{\infty} {\frac {{\rm d}N^{\rm line}_{\rm Comp}(r_{\rm
j};r_{\rm j}>r_{\rm j,tr})}{{\rm d}t{\rm d}r_{\rm j} }}{\rm
d}r_{\rm j},{\rm if}~ r_{\rm j,min}<r_{\rm j,tr};\\
\int\limits_{r_{\rm j,min}}^{\infty} {\frac {{\rm d}N^{\rm
line}_{\rm Comp}(r_{\rm j};r_{\rm j}>r_{\rm j,tr})}{{\rm d}t{\rm
d}r_{\rm j} }}{\rm d}r_{\rm j},{\rm if}~ r_{\rm j,min}\ge
r_{\rm j,tr}. \label{dnscatdt}
\end{cases}
\end{equation}

We can roughly estimate the luminosity of H$\alpha$ line emitted
from the disk region between $\xi_1$ and $\xi_2$ by
\begin{equation} L_{\rm H\alpha}\sim h\nu_{\rm H\alpha} {\frac
{{\rm d}N_{\rm Comp}^{\rm line}}{{\rm d}t}},\label{lhalpha}\end{equation} if the
line emitting portion is photo-ionized by the X-ray photons
scattered in the jet and the thermalization is not important.

We can derive $n_{\rm e}(r_{\rm j})$ from Eqs. (\ref{lkin1}), and (\ref{mdotjet}),
if the ratio $L_{\rm kin}/L_{\rm Edd}$ is specified. Using Eqs. (\ref{lkin1}),
(\ref{dnscat}), (\ref{dnscatthin}), (\ref{dnscatdt}) and (\ref{lhalpha}), we can calculate $L_{\rm H\alpha}/L_{\rm disk}$ as
a function of $L_{\rm kin}/L_{\rm Edd}$, when $\gamma_{\rm j}$,
$\gamma_{\rm e,min}$, and $\alpha_{\rm e}$ are specified.

\figurenum{2}
\centerline{\includegraphics[angle=0,width=8.0cm]{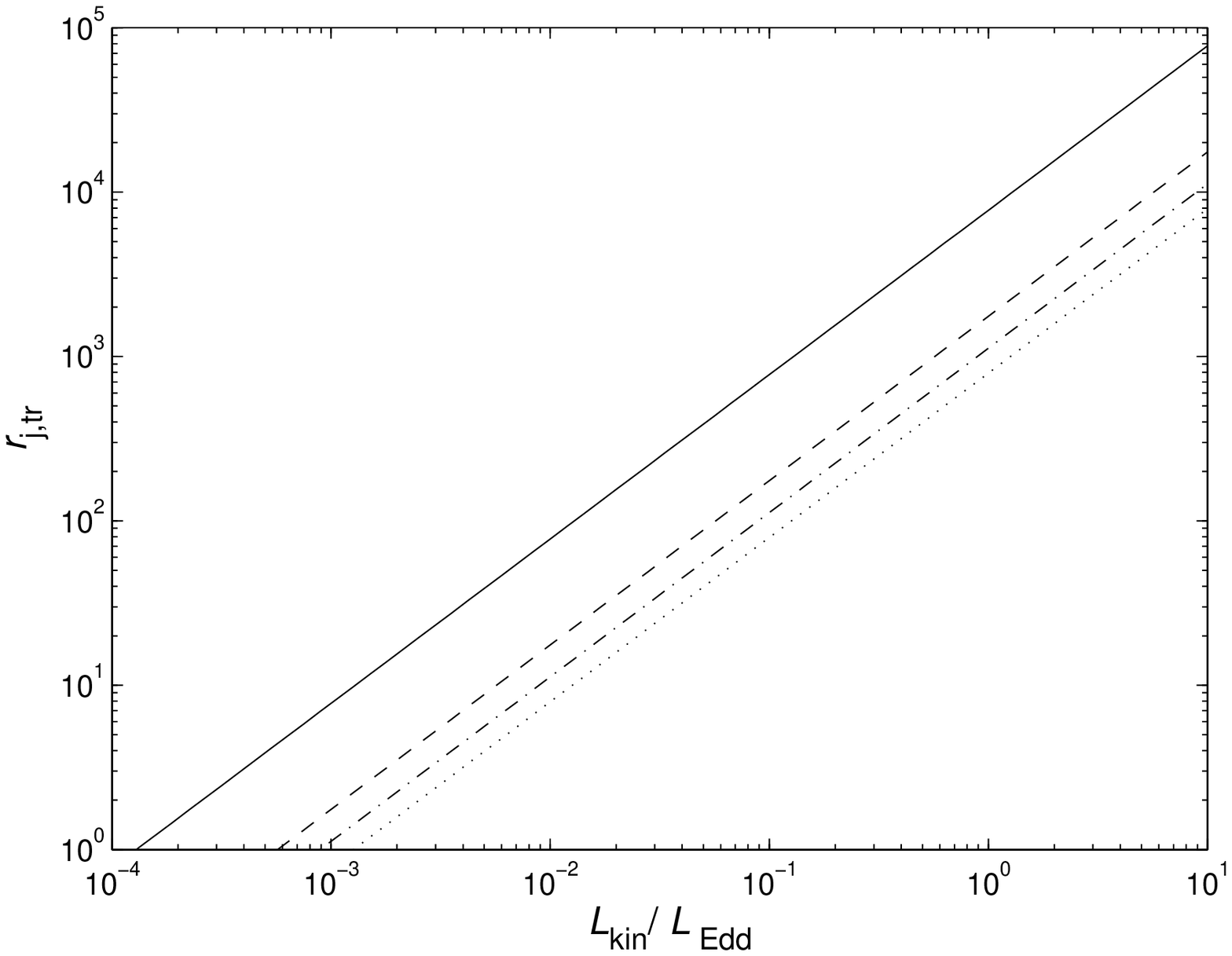}}
\figcaption{The radii $r_{\rm j,tr}$ of the
jets transit from Compton thick to Compton thin as functions of
$L_{\rm kin}/L_{\rm Edd}$ for different values of $\gamma_{\rm
j}=1.01$ (solid), 1.2 (dashed), 1.5 (dash-dotted), and 2 (dotted), respectively.
\label{fig2}}\centerline{}

\figurenum{3}
\centerline{\includegraphics[angle=0,width=8.0cm]{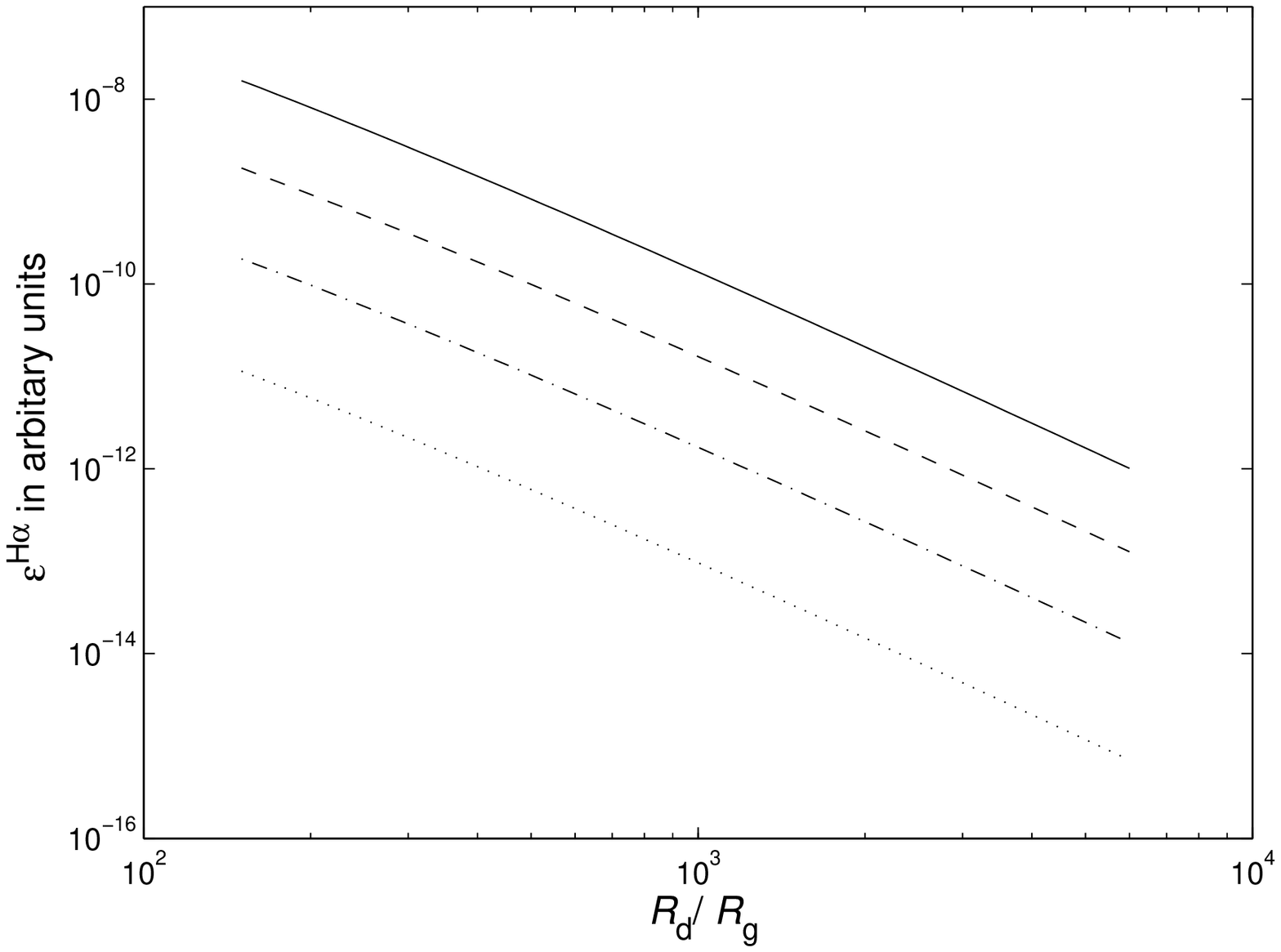}}
\figcaption{The emissivities $\epsilon^{{\rm
H}\alpha}$ as functions of disk radius $R_{\rm d}/R_{\rm g}$ for
different values of $\gamma_{\rm j}$=1.2 (solid), 1.5 (dashed),
2 (dash-doted) and 3 (doted), respectively. \label{fig3}}\centerline{}

\section{Testing the model}

In this paper, we use the sample of RL AGNs with double-peaked
broad lines given by \citet{eh03} to test our model. There are 20
sources with detected double-peaked H$_\alpha$ lines in this
sample (see Table 7 in their paper, and references therein). The
bolometric luminosity is derived from soft X-ray luminosity. They
found that the viscous power output $W_{\rm d}$ of the
line-emitting disk region is insufficient for the observed
H$_\alpha$ luminosity for these sources, of which seven sources
even have $W_{\rm d}<L_{{\rm H}\alpha}$. We list the bolometric
and H$_\alpha$ luminosities in Table 1 (converted to the cosmology
adopted in this paper).

We estimate the jet power of these sources from their
low-frequency radio emission. The relation between jet power and
radio luminosity proposed by \citet{w99} is
\begin{equation} Q_{\rm jet}\simeq 3\times 10^{45}f^{3/2}L_{\rm
151}^{6/7}~~ {\rm erg~s^{-1}}, \label{qjetrad}
\end{equation} where $L_{\rm 151}$ is the total radio luminosity at 151
MHz in units of 10$^{28}$ W~Hz$^{-1}$~sr$^{-1}$. \citet{w99} have
argued that the normalization is uncertain and introduced the
factor $f$ to account for these uncertainties. They use a wide
variety of arguments to suggest that  $1\leq f\leq 20$.
\citet{br00} argued that $f\sim 10$ is a likely consequence of the
evolution of magnetic field strengths as radio sources evolve. In
this paper, we adopt $f=10$ in all our calculations.

Finally, we need to estimate the masses of the black holes in
these double-peaked line emitters. It is still unclear whether the
empirical relation between the BLR size and optical continuum
luminosity suggested by \citet{k00} still holds for these AGNs
\citep*[Wang et al. 2005; but also see][]{wl04}, because the double-peaked lines are
assumed to originate from accretion disks, which is different from
the conventional cloud models for BLRs. In this paper, we use the
relation between host galaxy absolute magnitude $M_{\rm R}$ at
R-band and black hole mass $M_{\rm bh}$ proposed by \citet{md04},
\begin{equation}
 \log_{10}(M_{\rm bh}/{\rm M}_{\odot})=-0.50(\pm0.02)M_{\rm R}-2.75(\pm0.53),\label{mrmbh}
\end{equation} to estimate the black hole masses. There are seven sources in this
sample, of which the host galaxy magnitudes are unavailable in
literature. We adopt the total magnitudes (including the nuclear
emission), which may over-estimate the black hole masses for these
seven sources.

In this work, we adopt $\phi_{\rm j}=1/\gamma_{\rm j}$ for a free
jet \citep{bk79,hm86}, which is a good approximation even for a slow 
jet with $\gamma_{\rm j}\sim 1$. 
\citet{cf93} have argued that the lower energy cut for the
non-thermal distributed relativistic electrons in the jets can be
as low as $\gamma_{\rm e,min}\sim 1$. Our model calculations are
carried out for $\gamma_{\rm e,min}=1$. The
spectral index for optically thin synchrotron radiation
$\alpha_{\rm e}=0.75$ is adopted for all calculations in this
work, which is consistent with observations on radio-loud AGNs
\citep*[e.g.,][]{j98}. The model calculations can be carried out,
if the integral limits: $r_{\rm j,min}$, $\xi_1$, and $\xi_2$ are
specified. In all our calculations, $r_{\rm j,min}=10$ is adopted.
In Fig. 2, we show how the transition radius $r_{\rm j,tr}$ of the
jet from Compton thick to Compton thin varies with jet kinetic
luminosity $L_{\rm kin}$. The transition radius $r_{\rm j,tr}$
increases with $L_{\rm kin}$. The jet can be Compton thin down to
$r_{\rm j,min}$, when $L_{\rm kin}/L_{\rm Edd}\la
10^{-3}-10^{-2}$. For the cases with high kinetic luminosity
$L_{\rm kin}/L_{\rm Edd}$, the jet is Compton thick even at very
a large radius $r_{\rm j}$. We integrate Eq. (\ref{dnscatdt}) over
$r_{\rm j}$ (from $r_{\rm j,min}=10$) to calculate the
emissivities $\epsilon^{\rm H\alpha}$ as functions of radius in
Fig. \ref{fig3} for different values of $\gamma_{\rm j}$. The
final results depend insensitively on the value of $r_{\rm
j,min}$, because the solid angle of the outer line-emitting region
subtended to the jet is very small if $r_{\rm j}$ is small. It is
found that the emissivities have nearly power-law $r$-dependent distributions:
$\propto r_{\rm d}^{-\beta}$, and the values of $\beta$ are $\sim 2.5$.
This implies that the ratio $L_{\rm H\alpha}/L_{\rm disk}$ is
mainly governed by the inner radius $\xi_1$ of the line-emitting
disk region. We integrate Eq. (\ref{dnscatdt}), and the relations
between $L_{\rm kin}/L_{\rm Edd}$ and $L_{\rm H\alpha}/L_{\rm
disk}$ are plotted in Fig. \ref{fig4} for different values of
$\gamma_{\rm j}$. The
data of the sample are plotted in the same figure for comparison.

\figurenum{4}
\centerline{\includegraphics[angle=0,width=8.0cm]{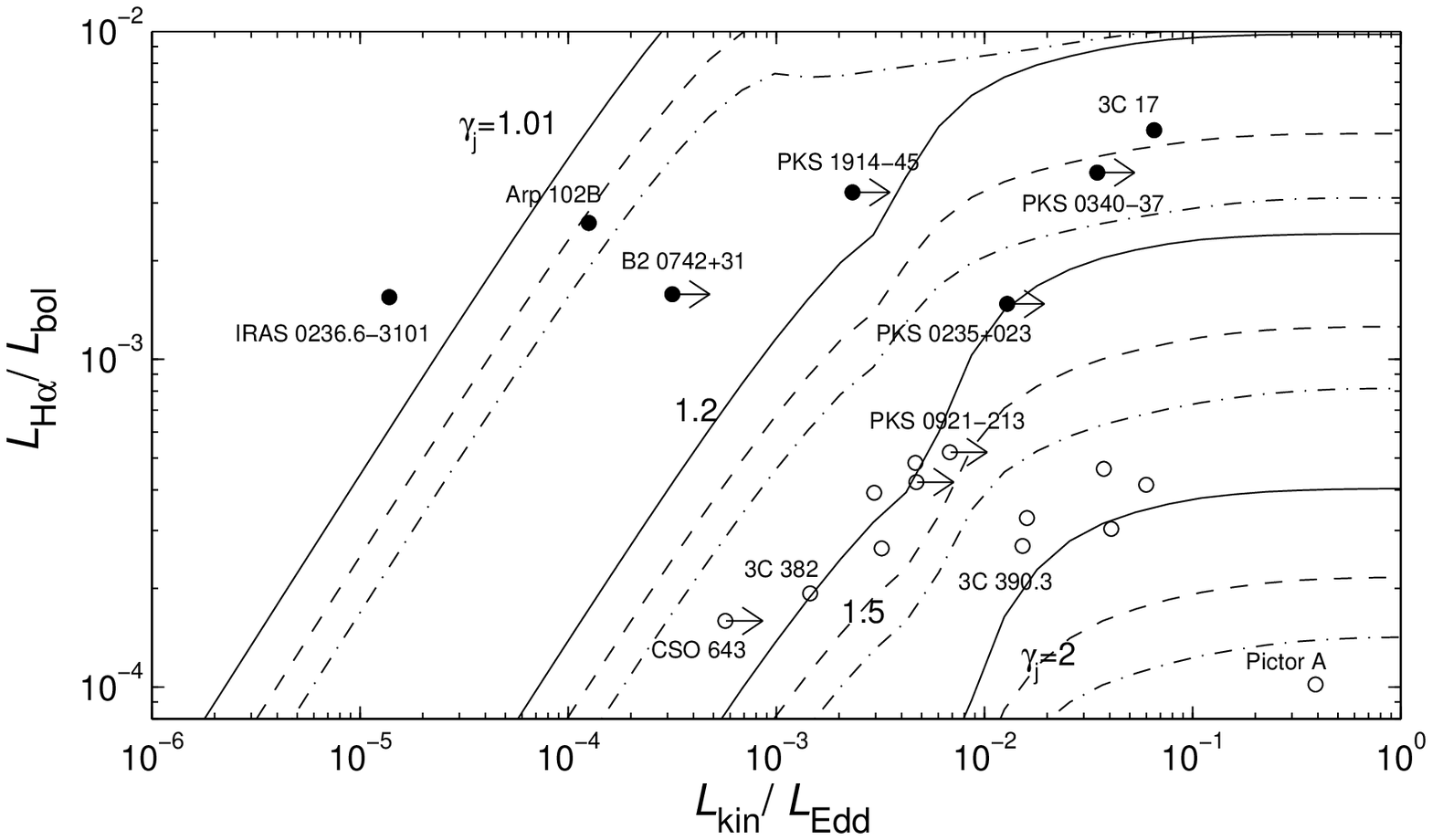}}
\figcaption{The relations between $L_{\rm
kin}/L_{\rm Edd}$ and $L_{{\rm H}\alpha}/L_{\rm bol}$
for different values of $\gamma_{\rm j}$. The lower electron
energy cutoff $\gamma_{\rm e,min}=1$ is adopted in the
calculations. The solid lines represent the line-emitting region
extending from $\xi_1=200$ to $\xi_2=3000$ (the dashed lines are
for $\xi_1=400$ and $\xi_2=3000$; the dash-dotted lines are for
$\xi_1=600$ and $\xi_2=3000$). The filled circles represent the
sources with $W_{\rm d}<L_{{\rm H}\alpha}$. \label{fig4}}\centerline{}

\section{Discussion}

In the ion-supported torus scenario, the torus is truncated to a
thin disk at a radius $R_{\rm d,tr}$, and the X-ray photons from
the torus are thought to illuminate the outer line-emitting disk
region. The most favorable candidate for such an ion-supported hot
torus is the RIAF \citep{ny94,ny95}. However, some double-peaked
emitters have very luminous X-ray emission, which cannot be
reproduced solely by the RIAFs in these sources if a reasonable
viscosity parameter is adopted (see discussion in \S 2). It is
quite doubtful whether RIAFs are in these X-ray luminous
double-peaked line emitters. In Fig. 1, we find that most energy
is radiated in the inner edge of the RIAF (more than 70 per cent
of the total radiation of the RIAF is emitted from the region
$R_{\rm d}\la 0.3 R_{\rm d,tr}$). Our calculation shows that less
than 2.3 per cent radiation from the RIAF in the inner region of
the disk can illuminate the outer disk region with $R_{\rm d}\ge
R_{\rm d,tr}$ (see \S 2), which implies that the radiation of
RIAFs is unable to solve the energy budget problem for most sources even if all the
observed luminous X-ray emission in some double-peaked line
emitters can be attributed to RIAFs.

Comparison between jet power estimated from the
extended radio emission and kinetic luminosity of the jet requires
$\gamma_{\rm e,min}\sim 1$ for electron-positron jets, and
$\gamma_{\rm e,min}\sim 100$ for electron-proton jets
\citep*[e.g.,][]{cf93}. For electron-proton jets, the
inverse-Compton scattered photons by the electrons in the jet are
in hard X-ray/$\gamma$-ray bands, which are inefficient for
photo-ionizing the line-emitting disk region. In our present
model, the line-emitting disk region is assumed to be irradiated
by the scattered X-ray photons from the electron-positron jet.
Most photons Compton up-scattered by the electrons in the
electron-positron jets are in the soft X-ray band, which can
efficiently photo-ionize the outer line-emitting disk region. Our
calculations show that the jet can be Compton thin down to $r_{\rm
j,tr}$ if the jet kinetic luminosity $L_{\rm kin}/L_{\rm Edd}$ is
low, while it can be Compton thick even at very large radii for
high values of $L_{\rm kin}/L_{\rm Edd}$, because much matter is
loaded by the jet with high $L_{\rm kin}/L_{\rm Edd}$. In our
calculations, the minimal radius of the jet $r_{\rm j,min}=10$ is
adopted. In principle, we can calculate the cases for a jet down
to a radius lower than $r_{\rm j,min}=10$, though it is still
unclear whether the K\"onigl's conical jet is still valid to a
very small radius near the black hole. Fortunately, the photons
scattered near the bottom of the jet cannot photo-ionize the outer
line-emitting disk region efficiently, because of the small solid
angle subtended to the line-emitting region, which is similar to
the RIAF case as discussed in \S 2. Thus, the final results have
not been affected much even if a value of $r_{\rm j,min}<10$ is
adopted.

In Fig. \ref{fig3}, we find that the emissivities $\epsilon^{{\rm
H}\alpha}$ have similar power-law $r$-dependence for different 
Lorentz factors $\gamma_{\rm j}$ adopted. We find $\epsilon^{{\rm
H}\alpha}$ varies with $r$ nearly $\propto r^{-2.5}$ along the disk radius for all
cases. The model fittings on the observed double-peaked line
profile requires a power-law line emissivity with an index of
$\sim 2-3$ for different sources \citep*[e.g.,][]{ch89,chf89,eh94,eh03}, which is consistent
with our model calculations. It is found that the emissivity
decreases rapidly with increasing the Lorentz factor $\gamma_{\rm
j}$. For given jet kinetic luminosity, a higher $\gamma_{\rm j}$
leads to lower electron density (see Eqs. \ref{lkin1} and \ref{mdotjet}), which
decreases the inverse-Compton scattered photons from the jet, and
then the emissivity $\epsilon^{{\rm H}\alpha}$.

The ratios $L_{{\rm H}\alpha}/L_{\rm disk}$ varying with $L_{\rm
kin}/L_{\rm Edd}$ for different values of $\gamma_{\rm j}$ are
plotted in Fig. \ref{fig4}. We plot the data of
the sample in Fig. \ref{fig4} to test our model by
comparison with the model calculations. We find that the Lorentz
factors of the jets $\gamma_{\rm j}\la 2$ are required to
explain the observed data for most sources in this sample. The apparent 
angular velocity of the brightest jet
component in 3C390.3 measured by VLBI is $\sim$0.54 mas/year \citep{k04}.
The Lorentz factor of the jet in this source $\gamma_{\rm j}\simeq 2.1$
is therefore derived from its inclination angle $i=26^\circ$,
which is estimated from the fitting of its double-peaked line
profile by \citet{eh94}. In Fig. \ref{fig4}, one
can find that a jet with $\gamma_{\rm j}\sim 1.5-2$ is required
by the model calculation for this source. For this source, $W_{\rm d}\simeq 4.6 L_{\rm H\alpha}$ \citep{eh03}, 
which means that the Blamer line emission may be partly attributed to the local 
viscous heating of the line-emitting disk region, so our 
result is roughly consistent with that derived from the VLBI observations. Only one source, Pictor
A, a jet with relatively larger Lorentz factor $\gamma_{\rm j}>2$
is required to illuminate the line-emitting region. This is a
powerful radio galaxy \citep*[e.g.,][]{w01}, and X-ray jet
emission is detected in this source \citep{h05}. The X-ray
emission from the jet is important for this source, and the
accretion disk luminosity derived from its X-ray luminosity is
obviously over-estimated, so the location of this source in Fig.
\ref{fig4} should be shifted upwards. A mild relativistic jet may
probably provide sufficient scattered photons for photo-ionizing
the line-emitting disk region in this source, if the X-ray
emission is dominantly from the jet. {For Arp 102B, a very low 
$\gamma_{\rm j}\sim 1.01$ ($\sim 0.14c$) is required (see Fig. \ref{fig4}), 
i.e., a slowly moving conical outflow 
is present in this source, which is consistent with its compact radio structure \citep{b81,p86,c01}. } Another special source, IRAS
0236.6-3101, which is difficult to be explained by this model (see
Fig. \ref{fig4}). This is a star-forming galaxy
with very weak radio emission (9.5 mJy at 1.4 GHz), and it
probably has no jet. The origin of its double-peaked lines may be
similar to that of radio-quiet double-peaked line emitters, which
will be addressed in the last paragraph of this section.

For the seven sources with only upper limits on black hole masses
(lower limits on the ratios $L_{\rm kin}/L_{\rm Edd}$), they are
similar to others and the jets with moderate Lorentz factors
$\gamma_{\rm j}\la 2$ can solve the ``energy budget" problem.
As the detailed model fittings for individual sources indicate
that the line-emitting regions in these sources have different
sizes \citep*[see][for details]{eh03}, our calculations have been
carried out for different values of $\xi_1$ and $\xi_2$(see
different line types in Fig. \ref{fig4}), which
have not altered our main conclusion.

There are 13 sources in this sample with $W_{\rm d}/L_{{\rm
H}\alpha}>1$. Although it means the local viscous power in the
line-emitting disk regions may be sufficient for the line emission
for these 13 sources, the temperatures of the line-emitting disk
portions are still too low for all the observed H$_\alpha$ line emission. The
external illuminating from the jets is still necessary for 
these sources. In our model, an electron-positron jet is required
to scatter disk photons back to photo-ionize the outer disk
surface to produce Balmer lines. We suggest that future VLBI
polarization observations on these radio-loud double-peaked line
emitters can test this model. 

In this paper, some quantities, $M_{\rm bh}$, $L_{\rm bol}$, and $L_{\rm kin}$, are 
derived from observed quantities by using the conventional approaches reported in literature, 
of which the uncertainties might affect our results. However, the qulititative conclusion 
that slow jets/outflows are required in double-peaked emitters would not be altered (see Fig. 4), 
for typical errors for these quantities, i.e., 
a factor of $\sim 3$ for $M_{\rm bh}$, and the derived kinetic luminosity $L_{\rm kin}$ 
is accurate at an order of magnitude. {In our present calculation of 
$L_{\rm H\alpha}$, we have not included the thermalization od the disk. If the 
thermalization is considered, the resulted $L_{\rm H\alpha}$ will become  
lower, so a lower bulk Lorentz factor $\gamma_{\rm j}$ of the jet 
than our present model calculation is required (see Fig. \ref{fig4}). }  

Besides radio-loud AGNs in this sample, several ten radio-quiet
double-peaked line emitters have been discovered \citep{s03}.
Although no jet is present in these radio-quiet AGNs, we speculate
that slow moving outflows may be in these sources, which play the
same role of the jets in their radio-loud counterparts. If the
electrons in the outflows are non-relativistic, a fraction of disk
power radiated from its inner region can be Thomson scattered to
the outer line-emitting portion of the disk. If the Thomson
scattering depth is $\sim 0.2$, the outflows can scatter nearly
one-tenth of the disk radiation to illuminate the outer
line-emitting disk portion. If this is the case, the absorption by
the outflows in these radio-quiet AGNs is expected to be detected
by X-ray observations, which can be a test on our model. The lack
of detected double-peaked broad emission lines in strong
radio-loud quasars with relativistic jets 
is a natural prediction of our model, i.e., the Compton scattering
power is too weak to irradiate the outer line-emitting portion for
fast moving jets. For radio-quiet AGNs, only those having outflows
with suitable scattering depth can scatter disk photons back to
illuminate the outer line-emitting portion, which leads to
double-peaked broad emission lines. Recent X-ray observations showed that 
a fraction of quasars and Seyfert galaxies have X-ray absorption with 
column density $\ga 10^{23} {\rm cm}^{-3}$ \citep*[e.g.,][]{row03,y05,p05}, 
which implies that the AGNs having outflows with suitable scattering depth $\sim 0.2$ 
are not common among all AGNs. {However, it does not mean that all AGNs with suitable 
scattering depth will exhibit double-peaked lines. Some other factors, such as, the viewing angle 
and the properties of the accretion disk, may play important roles in producing double-peaked lines.}  
The quantitative model calculations to fit the
line profiles for individual AGNs are beyond the scope of this
work, which will be reported in the forthcoming paper.

{In this work, we suggest that the soft photons scattered by electrons in electron-position jets for 
radio-loud double-peaked line emitters (or in slow outflows with suitable scattering depth for radio-quiet counterparts) can 
efficiently ionize the line-emitting disk regions. Another possibility is that slow outflows play the same 
role in radio-loud double-peaked line emitters as radio-quiet sources, though it is still unclear whether 
relativitic jets can co-exist with such cold outflows in radio-loud double-peaked line emitters. However, 
it is unable to explain the fact that no double-peaked
emission line is present in strong radio quasars with relativistic jets, while this is a natural prediction of the electron-positron jet model proposed in this work.}

\acknowledgments  We thank the referees for their helpful comments and suggestions. 
This work is supported by the National Science
Fund for Distinguished Young Scholars (grant 10325314), and the
NSFC (grant 10333020).

\begin{deluxetable}{ccccccccc}
\tabletypesize{\scriptsize} \tablecaption{Data of the sample }
\tablewidth{0pt} \tablehead{ \colhead{Source name} &
\colhead{redshift} & \colhead{$\log_{10} M_{\rm bh}/{\rm
M}_\odot$} & \colhead{reference} & \colhead{$\log_{10} L_{{\rm
H}\alpha}$} & \colhead{$\log_{10} L_{\rm bol}$} &
\colhead{$\log_{10} Q_{\rm jet}$} & \colhead{$L_{{\rm
H}\alpha}/L_{\rm bol}$} & \colhead{$Q_{\rm jet}/L_{\rm Edd}$} }
\startdata
 3C 17 &  0.220 & 8.81$^{\rm a}$ & 1 & 43.50 & 45.80 & 45.73 &  5.0$\times 10^{-3}$ & 5.92$\times 10^{-2}$ \\
 4C 31.06 &  0.373 & 9.66$^{\rm a}$ & 2 & 43.08 & 46.66 & 45.17 & 2.65$\times 10^{-4}$ &  2.91$\times 10^{-3}$\\
 3C 59 & 0.109 & 8.59 & 3 & 42.82 & 46.30 & 44.89 & 3.28$\times 10^{-4}$ & 1.45$\times 10^{-2}$ \\
 PKS 0235+023 & 0.209 & $<$8.91$^{\rm a}$ & 4 & 43.85 & 46.68 & 44.95 & 1.48$\times 10^{-3}$ &  7.87$\times 10^{-3}$\\
 IRAS 0236.6-3101 & 0.062 & 8.83 & 5 & 42.20 & 45.01 & 42.07 & 1.55$\times 10^{-3}$ &  1.25$\times 10^{-5}$\\
 PKS 0340-37 &  0.285 & $<$8.87 & 2 & 44.22 & 46.65 & 45.50 & 3.71$\times 10^{-3}$ & 3.16$\times 10^{-2}$\\
 3C 93 & 0.357 & 8.99 & 6 & 43.20 & 46.59 & 45.86 & 4.14$\times 10^{-4}$ & 5.43$\times 10^{-4}$\\
 MS 0450.3-1817 &  0.059 &  7.63$^{\rm b}$ & 7 & 41.28 & 44.69 & 43.20 & 3.91$\times 10^{-4}$ & 2.68$\times 10^{-3}$ \\
 Pictor A & 0.035 & 7.62$^{\rm a}$ & 8 & 41.77 & 45.76 & 45.31 & 1.02$\times 10^{-4}$ &  3.53$\times 10^{-1}$\\
 B2 0742+31 & 0.462 &  $<$11.22$^{\rm a}$ & 9 & 44.63 & 47.43 & 45.81 & 1.58$\times 10^{-3}$ &  2.87$\times 10^{-4}$\\
 CBS 74 &  0.092 &  $<$8.73$^{\rm a}$ & 2 & 42.92 & 46.29 & 44.50 & 4.22$\times 10^{-4}$ &  4.28$\times 10^{-3}$ \\
 PKS 0921-213 &  0.053 & $<$7.87$^{\rm a}$ & 2 & 42.42 & 45.70 & 43.80 & 5.21$\times 10^{-4}$ & 6.19$\times 10^{-3}$\\
 PKS 1020-103  & 0.197 & 8.85 & 10 & 43.54 & 46.86 & 44.61 & 4.83$\times 10^{-4}$ &  4.23$\times 10^{-3}$ \\
 CSO 643 & 0.276 & $<$9.77$^{\rm a}$ & 11 & 43.25 & 47.05 & 44.62 & 1.59$\times 10^{-4}$ &  5.16$\times 10^{-4}$ \\
 3C 303 & 0.141 & 8.27$^{\rm a}$ & 8 & 42.29 & 45.81 & 44.97 & 3.04$\times 10^{-4}$ & 3.269$\times 10^{-2}$ \\
 3C 332 & 0.151 & 8.53$^{\rm a}$ & 8 & 42.46 & 45.79 & 45.20 & 4.63$\times 10^{-4}$ &  3.40$\times 10^{-2}$\\
 Arp 102B & 0.024167 & 8.34 & 12 & 41.87 & 44.46 & 42.54 & 2.61$\times 10^{-3}$ & 1.14$\times 10^{-4}$ \\
 3C 382 & 0.05787 & 9.38$^{\rm a}$ & 1 & 43.07 &  46.78 & 44.64 & 1.93$\times 10^{-4}$ & 1.32$\times 10^{-3}$ \\
 3C 390.3 & 0.0561 & 8.59 & 13 & 42.57 & 46.14 & 44.87 & 2.69$\times 10^{-4}$ &  1.38$\times 10^{-2}$\\
 PKS 1914-45 & 0.368 & $<$10.37$^{\rm a}$ & 9 & 43.75 & 46.24 & 45.83 & 3.23$\times 10^{-3}$
 & 2.11$\times 10^{-3}$\\
 \enddata
 \tablenotetext{a}{The R-band magnitude is converted from V-band magnitude
 assuming $m_{\rm V}-m_{\rm R}=0.61$ \citep{fsi95}.}
 \tablenotetext{b}{The R-band magnitude is converted from K-band magnitude
 assuming $m_{\rm R}-m_{\rm K}=2.5$ \citep{d03}.}
 \tablecomments{Column (3) the black hole mass is estimated from the host galaxy R-band
 magnitude except the two sources: Arp 102B and 3C390.3; Columns (5)$-$(7) in units of erg s$^{-1}$: }
 \tablerefs{(1) \citet{sh89}; (2) \citet{eh03}; (3) \citet{md01};
 (4) \citet{s94}; (5) \citet{lv89}; (6) \citet{p03}; (7)
 \citet{2mass03}; (8) \citet{z96}; (9) NED; (10) \citet{f03}; (11)
 \citet{s01}; (12) \citet{f77}; (13) \citet{k00}. }
\end{deluxetable}



\end{document}